\documentstyle[aps,manuscript]{revtex}
\begin{document}
\title {Dynamic study on fusion reactions for $^{40,48}$Ca+$^{90,96}$Zr
 around Coulomb barrier
\footnote
{ *Email address: wangning@iris.ciae.ac.cn\\
 \dag Email address: lizwux@iris.ciae.ac.cn}}

\author {Ning Wang$^{1,*}$, Xizhen Wu$^{1,2,\dag}$,Zhuxia Li$^{1,2,3,\dag}$}

\address { 1) China Institute of Atomic Energy, P. O. Box 275(18), \\
            Beijing 102413, People Republic of China\\
           2) Nuclear Theory Center of National Laboratory of Heavy Ion
            Accelerator,\\Lanzhou 730000, People Republic of China\\
           3) Institute of Theoretical Physics, Chinese Academy of Sciences,\\
            Beijing 100080, People Republic of China }
\maketitle
\begin{abstract}
By using the updated improved Quantum Molecular Dynamics model in
which a surface-symmetry potential term has been introduced for
the first time, the excitation functions for fusion reactions of
$^{40,48}$Ca+$^{90,96}$Zr at energies around the Coulomb barrier
have been studied. The experimental data of the fusion cross
sections for $^{40}$Ca+$^{90,96}$Zr have been reproduced
remarkably well without introducing any new parameters. The fusion
cross sections for the neutron-rich fusion reactions of
$^{48}$Ca+$^{90,96}$Zr around the Coulomb barrier are predicted to
be enhanced compared with a non-neutron-rich fusion reaction. In
order to clarify the mechanism of the enhancement of the fusion
cross sections for neutron-rich nuclear fusions, we pay a great
attention to study the dynamic lowering of the Coulomb barrier
during a neck formation. The isospin effect on the barrier
lowering is investigated. It is interesting that the effect of the
projectile and target nuclear structure on fusion dynamics can be
revealed to a certain extent in our approach. The time evolution
of the N/Z ratio at the neck region has been firstly illustrated.
A large enhancement of the N/Z ratio at neck region for
neutron-rich nuclear fusion reactions is found.
\end{abstract}

\begin{center}
{\bf 1. INTRODUCTION}
\end{center}

Being encouraged by the synthesis of superheavy elements, the
investigation on fusion mechanism at low energies has recently
received a great attention both theoretically and experimentally
\cite{Zag01,Dia00,Ada00,Mar01,Ari00,Tro01,Tim98,Oga01}. Since the
central region of superheavy elements were predicted to locate at
Z=114 or 120 and N=184, which is strongly neutron-rich, the study
of dynamics for neutron-rich fusion reactions is highly demanded
for the purpose of the synthesis of superheavy elements. The
dynamics of fusion process for normal nuclear systems has been
studied in \cite{Bec85,Van92,Rei94,Bal98,Pro97,Hag97,Kon00}. In
these studies, it has been shown that the neck formation,
dynamical deformation, etc, result in a lowering of the fusion
barrier and furthermore it has been demonstrated that this
lowering effect is mostly significant at energies near the
barrier, consequently the sub-barrier fusion cross sections are
enhanced compared with the prediction of WKB approximation. But
for neutron-rich systems, the dynamics of fusion process is much
less studied. For neutron-rich systems, the symmetry term of EOS
should play a significant dynamical role. Therefore it seems to us
that it is highly requisite to study how the symmetry potential
influences the mechanism of neutron-rich fusion reaction process
dynamically. In this work, we devote ourselves to study the fusion
dynamics for neutron-rich systems at energies around the barrier
by means of the improved quantum molecular dynamics (ImQMD) model
\cite{Wang02}. In ref. \cite{Wang02} we have shown that the ImQMD
model can describe the properties of the ground state of selected
nuclei from $^{6}Li$ to $^{208}Pb$ very well with one set of
parameters and the experimental data of fusion reaction cross
sections for $^{40}$Ca+$^{90,96}$Zr \cite{Tim98} can also be
reproduced well with no extra-parameters. From that study, the
experimentally observed enhancement of fusion cross sections for
$^{40}$Ca+$^{96}$Zr compared with the non-neutron-rich fusion
reaction of $^{40}$Ca+$^{90}$Zr was attributed to gaining a
stronger dynamical lowering effect of the Coulomb barrier for the
neutron-rich target reaction of $^{40}$Ca+$^{96}$Zr. Based on that
investigation, it would be very interesting to study the dynamics
of fusion reactions induced by the neutron-rich projectile
$^{48}$Ca at energies around the Coulomb barrier with the same
model. As is well known that $^{48}$Ca has double closed shell
structure and spherical shape as the same as $^{40}$Ca. Therefore,
the static deformation effect of the projectile on the enhancement
of fusion cross sections at energies around the barrier can be
ruled out, and the role of the isospin effect should be shown up
by a comparison between two cases. But, on the other hand, the
shell structure of $^{48}$Ca is rather different from $^{40}$Ca
and the energy of octupole vibrations of $^{48}$Ca is about 1 MeV
higher than that of $^{40}$Ca due to the shell structure.
Furthermore,from the inelastic scattering study it has been shown
that $^{40}$Ca has a stronger octupole vibration than
$^{48}$Ca\cite{Tro01}. The situation is different for Zr isotopes
for which the energy of 3$^{-}$ state decreases as the number of
neutrons increases from $^{90}$Zr to $^{96}$Zr. This structure
effect should influence the fusion dynamics and the fusion cross
sections as well. It is not clear how to explicitly implement this
effect into our model at this moment. However, a dynamical study
on the neutron-rich fusion reactions can provide us with the
information about dynamical deformation which may relate to the
structure of projectile and target in addition to the information
of isospin effect on a fusion process, which are quite general. In
this work, we make comparison of the dynamic barrier lowering
effect for  4  reaction systems $^{40,48}$Ca+$^{90,96}$Zr at
energies around the barrier, and furthermore we analyze the causes
for the dynamic barrier lowering in detail, mainly focus on the
stage of the neck formation and neck development.

The paper is organized as follows: In sec. II we briefly introduce
our ImQMD model. Then we study the mechanism of neutron-rich
nuclear fusion reactions in sec. III. Finally, a short summary and
discussion are given in sec IV.

\begin{center}
\bigskip {\bf II. IMPROVED QMD MODEL}
\end{center}

For reader convenience, in this section we briefly introduce the
ImQMD model. In the ImQMD model as the same as in the original QMD
model\cite{Hart89,Ai91,Hart98,Ono92}, each nucleon is represented
by a coherent state of a Gaussian wave packet
\begin{equation}  \label{1}
\phi _{i}({\bf r})=\frac{1}{(2\pi \sigma _{r}^{2})^{3/4}}\exp [-\frac{%
({\bf r-r}_{i})^{2}}{4\sigma _{r}^{2}}+\frac{i}{\hbar}{\bf r}\cdot
{\bf p}_{i}],
\end{equation}
where ${\bf r}_{i}$ and ${\bf p}_{i}$ are the centers of the i-th
wave packet in the coordinate and momentum space, respectively.
$\sigma _{r}$ represents the spatial spread of the wave packet.
Through a Wigner transformation of the wave function, the one-body
phase space distribution function for N-distinguishable particles
is given by:
\begin{equation}  \label{2}
f({\bf r,p})=\sum\limits_{i}f_{i}({\bf r,p}).
\end{equation}
where
\begin{equation}  \label{3}
f_{i}({\bf r,p})=\frac{1}{(\pi\hbar)^{3}}\exp[-\frac{({\bf
r-r}_{i})^{2}}{2\sigma_{r}^{2}}-\frac{2\sigma_{r}^{2}}{\hbar^{2}}({\bf
p-p}_{i})^{2}].
\end{equation}

For identical Fermions, the effects of the Pauli principle are
discussed in a broader context by Feldmeier and Schnack
\cite{Fel00}. The approximative treatment of anti-symmetrisation
used in this paper is explained below. The density and momentum
distribution function of a system read
\begin{equation}  \label{4}
\rho ({\bf r})=\int f({\bf r,p})d^{3}p=\sum\limits_{i} \rho
_{i}({\bf r}),
\end{equation}

\begin{equation}  \label{5}
g({\bf p})=\int f({\bf r,p})d^{3}r=\sum\limits_{i} g_{i}({\bf p}),
\end{equation}
respectively, where the sum runs over all particles in the system.
$\rho _{i}({\bf r})$ and $g_{i}({\bf p})$ are the density and
momentum distribution function of nucleon i:
\begin{equation}  \label{6}
\rho _{i}({\bf r})=\frac{1}{(2\pi \sigma _{r}^{2})^{3/2}}\exp [-\frac{%
({\bf r-r}_{i})^{2}}{2\sigma _{r}^{2}}],
\end{equation}

\begin{equation}  \label{7}
g_{i}({\bf p})=\frac{1}{(2\pi \sigma _{p}^{2})^{3/2}}\exp [-\frac{%
({\bf p-p}_{i})^{2}}{2\sigma _{p}^{2}}],
\end{equation}
where $\sigma_{r}$ and $\sigma_{p}$ are the widths of wave packets
in coordinate and momentum space, respectively, and they satisfy
the minimum uncertainty relation. The time evolution of ${\bf r}_{i}$ and ${\bf
p}_{i}$ is governed by Hamiltonian equations of motion:
\begin{equation}  \label{8}
\dot{{\bf r}}_{i}=\frac{\partial H}{\partial {\bf p}_{i}},
\dot{{\bf p}}_{i}=-\frac{\partial H}{\partial {\bf r}_{i}}.
\end{equation}

The Hamiltonian H consists of the kinetic energy and the effective
interaction potential energy:
\begin{equation}  \label{9}
H=T+U.
\end{equation}

The effective interaction potential energy includes the nuclear
local interaction potential energy and Coulomb interaction
potential energy:
\begin{equation}  \label{10}
U=U_{loc}+U_{coul},
\end{equation}

and
\begin{equation}  \label{11}
U_{loc}=\int V_{loc}d^{3}{\bf r}.
\end{equation}
$V_{loc}$ is the potential energy density, which can be derived
directly from a zero-range Skyrme interaction \cite{Dav88,Vau72}. Thus,

\begin{equation}  \label{12}
U_{loc}=\frac{\alpha }{2}\sum\limits_{i} \langle \frac{\rho }{\rho _{0}}\rangle _{i}+%
\frac{\beta }{3}\sum\limits_{i} \langle \frac{\rho}{\rho _{0}}\rangle ^{2}_{i}+%
\frac{C_{s}}{2}\int \frac{(\rho _{p}-\rho _{n})^{2}}{\rho
_{0}}d^{3}{\bf r}+\int \frac{g_{1}}{2}(\nabla \rho )^{2}d^{3}{\bf
r},
\end{equation}
where
\begin{equation}
\langle \rho \rangle _{i}=\sum\limits_{j\neq i}\rho_{ij},
\end{equation}
and
\begin{equation}
\rho_{ij}=\frac{1}{(4\pi \sigma_{r}^{2})^{3/2}}\exp [-\frac{ ({\bf
r}_{i}-{\bf r}_{j})^{2}}{4\sigma _{r}^{2}}].
\end{equation}
The third term in the right hand side of (12) is the symmetry
potential energy. The gradient term in $U_{loc}$ is to account for
the surface energy and the correction to the second term in
Equ.(12)\cite{Wang02,Dav88}.

Because in this work we are going to study the isospin effect on
the fusion dynamics in neutron-rich nuclear fusion reactions we
pay a special attention to the symmetry potential term. Therefore
we make a more careful treatment on the symmetry potential term,
namely, in addition to the volume symmetry potential term, we
further introduce a surface symmetry potential term according to
the finite-range Liquid-Drop Model\cite{Mol95},which reads as

\begin{equation} \label{12}
U_{sur-sym}=\frac{C_{s} C_{k}}{2\rho_{0}} \sum\limits_{i,j\neq i}
s_{i} s_{j} \rho_{ij} \nabla_{i}^{2} \rho_{ij}.
\end{equation}

Where, $s_{i}$ is +1 for proton and -1 for neutron and $C_{k}$ is
the strength parameter for the surface symmetry term. We find this
term plays an important dynamical role for reactions
$^{48}$Ca+$^{90,96}$Zr but a minor role for
$^{40}$Ca+$^{90,96}$Zr. It reduces the fusion cross sections for
$^{48}$Ca+$^{90,96}$Zr considerably but almost does not change the
cross sections of $^{40}$Ca+$^{90,96}$Zr. The discussion about the
effect of this term will be given elsewhere. The parameters used
in this work are listed in Table I

Considering the fact that for a finite system the nucleons are
localized in a finite region corresponding to the size of the
system, the width of wave packets representing nucleons in the
system should have a relation with the size of the system. As the
same as in \cite{Wang02}, here we also adopt a system size
dependent wave packet width to account for the fact, that is,
\begin{equation}\label{16}
\sigma_{r}=0.16N^{1/3}+0.49,
\end{equation}
where $N$ is the number of nucleons bound in the system.

In order to overcome the difficulty in describing the Fermionic
nature of N-body system in the QMD model, an approximative
treatment of antisymmetrization is adopted, namely, we implement
the phase space constraint of the CoMD model proposed by
Papa.et.al.\cite{Pap01} into the model. It is requested by the
constraint that the one body occupation number in a volume $h^{3}$
of phase space centered at $({\bf r_{i},p_{i}})$ corresponding to
the centroid of wave packet of particle i should always be not
larger than 1 according to the Pauli principle. The one body
occupation number is calculated by
\begin{equation}  \label{17}
f^{ocu}_{i}=\sum\limits_{j}\delta_{\tau_{i}\tau_{j}}\delta_{s_{i},s_{j}}\int_{h^{3}}
f_{j}({\bf r,p})d^{3}{\bf r} d^{3}{\bf p},
\end{equation}
where $s_{i}$ and $\tau_{i}$ are the third component of spin and
isospin of particle i. We have made a check for time evolution of
individual nuclei from light nuclei to heavy nuclei and we found
that by taking the procedure of phase space constraint, the
requirement is reasonably satisfied and the phase space
distribution is prevented efficiently from evolving to be a
classical distribution from the initial nuclear ground state
distribution for a long enough time.

Concerning the collision part, an isospin dependent
nucleon-nucleon scattering cross section and Pauli-blocking are
used \cite{Li01,Rei94}. This part actually plays a minor role in a
fusion reaction.

In this work the initial density distribution of projectile and
target is obtained by Skyrme HF calculations\cite{Ham96,Ham99}.
The other procedures are the same as in \cite{Wang02}. The model
has been carefully checked and it turns out that the ImQMD model
works well in describing the ground state properties for nuclei
from $^{6}$Li to $^{208}$Pb and calculating the static Coulomb
barrier for fusion reactions as well as fusion cross sections for
$^{40}$Ca+$^{90,96}$Zr.

\begin{center}
{\bf 3. RESULTS}
\end{center}
Before coming to the numerical results for fusion reactions
$^{40,48}$Ca+$^{90,96}$Zr, let us first make a survey on the
configurations along a fusion path. In Fig.1 we illustrate one
typical fusion event of the head on reaction of
$^{40}$Ca+$^{90}$Zr at the energy 5 MeV below the barrier. In the
figure, we plot the dynamical barrier $V_{b}$ as a function of the
distance between the center of mass of projectile and that of
target. We will discuss the dynamical barrier in more detail in
the following section (section B) and the definition of it will be
given there. Simultaneously in sub-figures we plot the contour
plots of density distributions as well as the corresponding
single-particle potentials at 3 typical time, i.e. before, at, and
after reaching the highest value of the dynamic barrier along the
fusion path. The single-particle potential is calculated by
\begin{equation}  \label{22}
V_{sp}({\bf r})=\int \rho ({\bf r}^{\prime })V({\bf r-r}^{\prime
})d^{3}{\bf r}^{\prime },
\end{equation}
with $\rho({\bf r})$ being the density distribution of the system
and $V({\bf r-r}')$ the effective nucleon-nucleon interaction. In
sub-figures(1a)and (1b) we plot the contour plot of the density
distribution as well as the corresponding single-particle
potential at the point 1 along the fusion path. One can find from
these two sub-figures that at this point the fusion partners are
not in touch( see sub-figure (1a)) and there is a high enough
inner potential barrier which prevents nucleons moving from the
projectile to target or vice versa ( see sub-figure(1b)). At the
time corresponding to the point 2, the dynamic barrier reaches a
maximum value. The contour plot of density distribution(
sub-figure(2a)) shows that the fusion partners are at the touching
configuration and a neck starts to grow and following it, the
inner potential barrier in the potential well is reduced allowing
a few nucleons moving from projectile to target or vice versa(see
sub-figure(2b)). At the time corresponding to the point 3, the
dynamical barrier is reduced considerably. Sub-fig.(3a) and
sub-fig.(3b) show that the neck develops considerably at this time
and consequently, the inner potential barrier in the potential
well is reduced substantially and nucleon transfer between the
projectile and target becomes much easier than before. This means
that a pre-compound nucleus begins to be formed. From this study
we have learned that how the dynamical fusion barrier is
correlated with the development of the configuration of fusion
partners along the fusion path.

In the following, we show the numerical results for fusion
reactions $^{40,48}$Ca+$^{90,96}$Zr. First we show the fusion
cross sections. For understanding the mechanism of the enhancement
of the fusion cross sections for $^{40}$Ca+$^{96}$Zr and
$^{48}$Ca+$^{90,96}$Zr compared with $^{40}$Ca+$^{90}$Zr case, we
show the dynamic barrier and the other quantities relevant to the
dynamic lowering of the Coulomb barrier only at head on reactions.
Following it we make a discussion about the isospin and structure
effect in fusion dynamics for the systems studied. In order to
exploring how the isospin transfers at the neck region, we study
the time evolution of the N/Z ratio at the neck region for
$^{40,48}$Ca+$^{90,96}$Zr reactions to see how it depends on the
initial N/Z ratio.

{\bf A. FUSION CROSS SECTIONS FOR $^{40,48}$Ca+$^{90,96}$Zr}

After making the preparation of initial nuclei, we elaborately
select ten projectile nuclei and ten target nuclei from thousands
of pre-prepared systems. By rotating these prepared projectile and
target nuclei around their centers of mass by a Euler angle chosen
randomly, we create 100 bombarding events for each reaction energy
$E$ and impact parameter $b$. Through counting the number of
fusion events, we obtain the probability of fusion reaction
$g_{fus}(E,b)$, then the cross section is calculated by using the
expression:
\begin{equation}  \label{19}
\sigma _{fus}=2\pi \int\limits_{0}^{b_{\max }}bg_{fus}(E,b)db=2\pi
\sum bg_{fus}(E,b)\Delta b.
\end{equation}
The distance from projectile to target at initial time is taken to
be 20 fm.

As for the definition of fusion event, we still adopt an
operational definition as the same as in TDHF calculations and in
the QMD model calculations\cite{Maru98}. More specifically, in
this work we consider any event, for which the number of nucleons
escaped during the process of forming compound nuclei is equal to
or less than 6, as a fusion event\cite{Wang02}.
 Fig.2 shows the fusion cross sections for
$^{40}$Ca+$^{90}$Zr, $^{40}$Ca+$^{96}$Zr,$^{40}$Ca+$^{90}$Zr, and
$^{40}$Ca+$^{90}$Zr, respectively. Experimental data for the
reactions of $^{40}$Ca+$^{90,96}$Zr taken from ref.\cite{Tim98}
are also shown. One can see that the experimental data for
$^{40}$Ca+$^{90,96}$Zr are reproduced well without introducing new
parameters and there is a strong enhancement of the fusion cross
sections for neutron-rich reactions. The fusion cross sections for
reactions $^{48}$Ca+$^{90,96}$Zr at energies around the barrier
are higher than those for $^{40}$Ca+$^{96}$Zr. But the enhancement
of the fusion cross sections for $^{48}$Ca+$^{90,96}$Zr compared
with $^{40}$Ca+$^{96}$Zr is not so strong as the case of
$^{40}$Ca+$^{96}$Zr compared with $^{40}$Ca+$^{90}$Zr. For
understanding the feature of the fusion excitation functions for
different systems shown in Fig.2, let us first look at the
distribution of fusion probabilities with respect to the impact
parameters in Fig.3. One can find in the figure that for
neutron-rich reactions, in addition to having a larger fusion
probability, the maximum impact parameter leading to fusion is
larger compared with non-neutron-rich reactions. For example, at
the incident energy 5 MeV below the static Coulomb barrier, the
maximum impact parameter leading to fusion is about 9 fm for
reaction $^ {48}$Ca+$^{90,96}$Zr , about 8.5 fm for $^
{40}$Ca+$^{96}$Zr, and only 6.5 fm for the non-neutron-rich
reaction of $^{40}$Ca+$^{90}$Zr. This means that for the
neutron-rich reactions, the fusion partners can be fused at a
relative larger distance. A possible reason for it is that the
dynamical elongation is enhanced for neutron-rich fusion systems.
The effect of the dynamical elongation on dynamical lowering of
the Coulomb barrier will be discussed in the following section.

For the cases of the incident energy at 10 MeV above the static
Coulomb barrier, the distribution of the fusion probability with
respect to the impact parameter shows a similar tendency but the
effect is weaker.

{\bf B. DYNAMIC LOWERING OF THE BARRIER}

In order to understand the reason for the enhancement of fusion
reaction cross sections for neutron-rich nuclear fusions in this
section we study the dynamic Coulomb barrier lowering effect. In
the QMD model, the Coulomb barrier is calculated microscopically
by using the following expressions
\begin{eqnarray}
V_{b}(d) & = & \int d^{3}r_{1}\int d^{3}r_{2}\rho _{1}({\bf
r}_{1}-{\bf r}_{1c})V({\bf r}_{1}-{\bf r}_{2})\rho _{2}({\bf
r}_{2}-{\bf
r}_{2c}),\\
d & = & |{\bf r}_{1c}-{\bf r}_{2c}|,\nonumber
\end{eqnarray}
where $\rho _{1}$, $\rho _{2}$ are the density distribution of
projectile and target, respectively; ${\bf r}_{1c}$, ${\bf
r}_{2c}$ are their centers of mass, respectively. $V({\bf r-r}')$
is the effective nucleon-nucleon interaction. It is clear that, in
general, $V_{b}(d)$ is a function of time since $\rho _{1}$, $\rho
_{2}$ changes from time to time. Only in a static case, where the
density distribution of projectile and target assumes to be the
same as that at the initial time and correspondingly the static
barrier is calculated with the static density distribution.
Therefore for the static barrier, the dynamical effects
experienced by fusion partners during reaction process are not
taken into account. For dynamic case, the density distribution of
the projectile and target is calculated by using expression(4)
with sum running over all particles in projectile and target,
respectively. When two colliding partners approach with each
other, the density distribution of projectile and target changes
from time to time and their shape (determined by the density
distribution) get deformed due to the interaction between them.
The time evolution of the shape deformation and the neck formation
depends on the incident system and energy as well as the impact
parameter. Consequently, the dynamical barrier not only depends on
the incident system but also depends on the incident energy as
well as the impact parameter. In the following we only study the
head on collision case and define the height of the highest
Coulomb barrier experienced in the path of fusion as the height of
the dynamic Coulomb barrier.

Generally, the dynamic barrier is lower than that of the static
one because of the neck formation and the increase of the N/Z
ratio at neck region for neutron-rich nuclear fusion reactions. As
an example, in table II we show the results about the dynamic
barrier for head on fusion reactions of $^{40,48}$Ca+$^{90,96}$Zr
at energies 5 MeV below and 10 MeV above the static Coulomb
barrier. From this table one can see that the dynamic effect
lowers the height of barrier dramatically and this dynamic
lowering is incident energy and system dependent. The barrier
lowering is stronger for the case of the energy below the barrier
than that of the energy above the barrier. This feature of barrier
lowering was also observed in \cite{Kon00} for symmetric reactions
of oxygen and nickel isotopes by means of the mean field transport
theory.

To illustrate the system dependence of the dynamic barrier, in
Fig.4 we show the time evolution of the dynamic barrier for head
on collisions of $^{40,48}$Ca+$^{90,96}$Zr at the incident energy
5 MeV below the static barrier. From a comparison among four
curves we see the following trends: 1) The dynamic Coulomb barrier
for neutron-rich reactions is lower than that for non-neutron-rich
reactions. 2) The barrier top position for neutron-rich reactions
is shifted to a larger distance compared to non-neutron-rich ones.
3) The width of the barrier for neutron-rich reactions is thinner
than that for non-neutron-rich reactions. As for three
neutron-rich reactions, there is no obvious difference in the
dynamic Coulomb barrier.

To investigate the causes leading to these trends let's turn to
study the quantities relevant to the dynamic barrier. For the
purpose of understanding the mechanism, in Table III we only give
the calculation results for head on collisions of
$^{40,48}$Ca+$^{90,96}$Zr at energies of 5 MeV below (lower energy
case) and 10 MeV above(higher energy case) the corresponding
static Coulomb barrier. The quantities listed in table III are
calculated as follows: for each event, we calculate the $V_{b}(d)$
at each time step to find out the time of reaching the highest
barrier( i.e. the point 2 in Fig. 1). The contour map with
$\rho=0.02/fm^{3}$ of the density distribution of the system at
this time gives the shape of the system (see sub-figure(2a) of
Fig.1). The schematic figure of Fig.5 illustrates the shape of the
system (typically for a head on collision) at this time and the
geometry quantities listed in table III, such as the distance
between the centers of mass of projectile and target, and the neck
radius,etc, are shown in the figure. The results given in Table
III are the average values of all corresponding events. The
elongation given in table III equals the distance between the
centers of mass of projectile and target minus the radii of
initial projectile and target nuclei. From Table III one can find
that the height of the dynamic barrier is closely correlated with
the elongations obtained for different incident energies and
collision systems listed in the table , i.e. the larger the
elongation is the lower the barrier is. Generally speaking, the
elongation at the touching configuration should depend on the
interaction time before reaching the touching configuration and
the longer interaction time leads to a larger elongation.
Therefore the elongation for the lower energy case is always
larger than that for the higher energy case. Table III shows that
the elongation for energy below static barrier case is about 10
$\%$ larger than that for above static barrier case. Furthermore,
the elongation also depend on structure of projectile and target
and the N/Z ratio at neck region as well. Now let's look at the
dependence of the elongation on the structure of reaction systems.
For the lower energy case, the largest elongation is obtained in
the reaction of $^{40}$Ca+$^{96}$Zr, while for the higher energy
case the largest elongation is obtained in $^{48}$Ca+$^{96}$Zr. As
is well known that the energy of octupole vibration of $^{96}$Zr
is lower than that of $^{90}$Zr and we may consider $^{96}$Zr is
softer than $^{90}$Zr. And for $^{48}$Ca, the energy of octupole
vibration is about one MeV higher than that of $^{40}$Ca , which
implies that $^{48}$Ca is more rigid than $^{40}$Ca. The
dependence of the elongation on the different systems given in
Table III clearly shows the influence of nuclear structure effect.
Concerning the isospin effect, it is quite natural that the
increase of N/Z at neck region should decrease the height of the
Coulomb barrier. There is a strong enhancement of the N/Z ratio at
neck region for neutron-rich reactions as shown in Table III.
Consequently, for $^{40}$Ca+$^{96}$Zr compared with
$^{40}$Ca+$^{90}$Zr both the isospin effect and the structure
effect are in favor to enhance the fusion cross sections of
$^{40}$Ca+$^{96}$Zr. While for reactions induced by $^{48}$Ca
compared with reactions induced by $^{40}$Ca, the isospin effect
and the structure effect are counterpart, consequently the
enhancement of fusion cross section induced by neutron-rich effect
is reduced by the structure effect.

{\bf C. TIME EVOLUTION OF THE N/Z RATIO AT THE NECK REGION}

As is seen from above study that the dynamic lowering of the
barrier is closely related to the configuration and component of
the neck. The N/Z ratio at the neck region is one of the most
sensitive quantities with respect to the neck formation for
neutron-rich nuclear fusion reactions, as shown in table III. For
the isospin symmetry case of $^{40}$Ca+$^{90}$Zr, the N/Z ratio at
the neck region is more or less the same as the average N/Z ratio
of the total system. But for the neutron-rich reactions, the N/Z
ratio at the neck region is much higher than that of the average
N/Z value of the corresponding systems. This effect results from
the different behavior of the density dependence of chemical
potential for neutrons and protons in isospin asymmetry systems.
The chemical potential is defined as
\begin{equation}
\mu_{n/p}=\frac{\partial
\varepsilon(\rho,\delta)}{\partial\rho_{n/p}},
 \end{equation}
where $\varepsilon(\rho,\delta)$ is the energy density and
$\mu_{n/p}$ and $\rho_{n/p}$ are the chemical potential and the
density of neutrons and protons,respectively. From the definition
one can find that the chemical potential is a function of both
density $\rho$ and isospin asymmetry $\delta$. Fig.6 shows the
chemical potential of proton and neutron as function of density
with $\delta=\frac{N-Z}{N+Z}=0.10$. From Fig.6 one can see that
the density corresponding to the minimum of the chemical potential
of neutrons is lower than that of protons for a neutron-rich
nuclear system and thus the neutrons are preferably drived to the
lower density area. This effect has also been studied and
confirmed in the intermediate energy heavy ion collisions. The
increasing of the N/Z ratio at the neck region should reduce the
dynamic barrier in fusion process. It would be interesting to
study the isospin transfer at neck region, therefore in Fig.7 we
show the time evolution of the N/Z ratio at the neck region for
head on fusion reactions of $^{40,48}$Ca+$^{90,96}$Zr at energies
5 MeV below and 10 MeV above the static Coulomb barrier, in which
the time is started from the beginning of neck formation(when the
density at the touching point reaches 0.02 $\rho_{0}$). The
general trend of the time evolution of the N/Z ratio is: the N/Z
ratio at the neck region first increases as time increases, then
soon reaches a maximum value and then decreases, finally it
approaches the average N/Z value of the system. The figure shows
that the enhancement of N/Z at neck region at the early stage of
the neck formation strongly depends on the N/Z ratio of the
initial system, i.e. the larger the isospin asymmetry of the
initial system is, the stronger the enhancement of N/Z ratio at
neck region is. The reason for the fluctuation appeared in the
time evolution of the N/Z ratio for neutron-rich reactions may be
understood as: at the beginning when the neck is just formed,
neutrons are preferably move to the neck region driven by the
chemical potential; not soon as too many neutrons are concentrated
there, the symmetry potential attracts more protons to migrate
into the neck region and the N/Z ratio is reduced, and then
because of the increase of proton number the Coulomb repulsion
plays a role.... Thus the interplay of the Coulomb force and the
symmetry potential results in the fluctuation behavior in time
evolution of the N/Z ratio at neck region for neutron-rich
systems. This fluctuation becomes stronger for non-central
collisions. With the growing of neck, nucleon transfer through the
neck becomes easier and the fusion system passes over the dynamic
barrier. After about 100 fm/c, that is, when a neck develops well,
the N/Z ratio at the neck region gradually approaches to the
average N/Z ratio of the whole system and the isospin degree of
freedom seems to gradually reach an equilibrium, but the
dissipation of the collective motion is still going on. The
details of the nucleon transfer and the dissipation of the
collective motion in the neck region will be discussed elsewhere.

\begin{center}
{\bf SUMMARY AND DISCUSSION}
\end{center}
In this work we have introduced a surface-symmetry potential term
into a QMD type transport model for the first time. We have used
this newly updated ImQMD model to study the fusion dynamics of
$^{40,48}$Ca+$^{90,96}$Zr at energies around the barrier. The
surface-symmetry term seems to play an important role in fusion
dynamics for $^{48}$Ca+$^{90,96}$Zr but negligible role in that of
$^{40}$Ca+$^{90,96}$Zr. Our calculated results of excitation
functions for fusion reactions of $^{40,48}$Ca+$^{90,96}$Zr show a
strong enhancement of fusion cross sections for the neutron-rich
reactions at energies near and below the static barrier. We have
made systematic analysis to understand this feature. We have shown
that the maximum impact parameter leading to fusion reaction for
neutron-rich reactions is larger than that for non-neutron-rich
reactions, which means that the excess neutrons make the reaction
partners to be fused at longer distance.

We have paid a great attention to study the dynamical fusion
barrier and found that there is a substantially lowering of the
dynamic barrier compared with the static Coulomb barrier due to
the neck formation. For the reactions studied we have observed
that: 1) The dynamic Coulomb barrier for a neutron-rich
configuration is lower than that for a non-neutron-rich case; 2)
The barrier top position for a neutron-rich configuration is
shifted to a larger distance compared to a non-neutron-rich
configuration; 3) The width of the barrier for a neutron-rich
configuration is thinner than that for a non-neutron-rich case.

We have shown that the time evolution of the ratio of neutrons to
protons ( the N/Z ratio ) at the neck region strongly depends on
the projectile and target isospin. At the early stage of the neck
formation, the N/Z ratio at neck region can reach a value of twice
the average N/Z ratio value of the whole system for
$^{48}$Ca+$^{90,96}$Zr, then after 100 fm/c later the N/Z ratio at
the neck region gradually approaches the average value of the
whole system, which means that the isospin degree of freedom
gradually approaches an equilibrium before the dissipation of
collective motion is completed.

A strong enhancement of fusion cross sections for
$^{40}$Ca+$^{96}$Zr compared to $^{40}$Ca+$^{90}$Zr have been
found which is in good agreement with the observation in
experiments. Relatively, the enhancement of fusion cross sections
for $^{48}$Ca+$^{90,96}$Zr compared with $^{40}$Ca+$^{96}$Zr is
less strong. The dynamic barrier lowering have been studied
systematically. We find it strongly relate to the elongation of
systems and the N/Z ratio at neck region at the touching
configuration on the fusion path. The results seem to show that
the elongation at touching configuration for different reaction
systems is correlated with the structure of projectile and target.
For instance, the largest elongation is obtained in the case of
$^{40}$Ca+$^{96}$Zr at 5 MeV below the static barrier in
consistent with the fact that the energy of octupole vibration
decreases from $^{90}Zr$ to $^{96}Zr$ and from $^{48}$Ca to
$^{40}$Ca as well. On the other hand, the isospin effect which
strongly influences the N/Z ratio at neck region for neutron-rich
nuclear fusion should affect the dynamic barrier strongly and
consequently, affect the fusion cross sections of neutron-rich
nuclear reactions. Further work on exploring how the isospin
effect and structure effect competes in fusion reactions is
needed. We strongly urge to make measurements of fusion cross
section and the distribution of barrier for
$^{40,48}$Ca+$^{90,96}$Zr to explore the interplay between these
two effects in fusion reactions.

The problem concerning the mass transfer is not discussed yet and
the neck dynamics is still not discussed thoroughly in this paper.
The work about these aspects is in progress.
\\

\begin{center}
{\bf ACKNOWLEDGMENTS}
\end{center}
We thank Profs. H.Q.Zhang and Z.H.Liu for stimulating discussions.
The work is supported by the National Natural Science Foundation
of China under Grant Nos. 19975073, 10175093, and 10175089, and by
the Science Foundation of Chinese Nuclear Industry and Major State
Basic Research Development Program under Contract No. G20000774.

\newpage

\begin{center}
{\small {\bf CAPTIONS} }
\end{center}

\begin{description}

\item[{\tt Fig.1}] The fusion path for a typical event of head on reaction of $^{40}$Ca+$^{90}$Zr
at the energy 5 MeV below the Coulomb barrier. The thick curve is
the dynamical barrier as a function of the distance between the
centers of mass of  projectile and target. Sub-figures (1a), (2a),
(3a) are for contour plots of the density distributions of the
reaction systems at the corresponding time pointed in the curve of
V$_{b} \sim$ d, and (1b), (2b), (3b)are the corresponding
single-particle potentials at the same time as Sub-figures (1a),
(2a), (3a).
\item[{\tt Fig.2}] The fusion cross sections for $^{40,48}$Ca+$^{90,96}$Zr.
The experimental data are taken from\cite{Tim98}.
\item[{\tt Fig.3}] The distributions of the fusion probability for reactions
of $^{40,48}$Ca+$^{90,96}$Zr with respect to impact parameters.
\item[{\tt Fig.4}] The dynamic barriers as a function of the distance of centers of mass projectile and
target for head on collisions of $^{40,48}$Ca+$^{90,96}$Zr at the
incident energy of 5 MeV below the static Coulomb barrier.
\item[{\tt Fig.5}] The definition of the geometric quantities in
Table III.
\item[{\tt Fig.6}] The density dependence of the chemical potential of protons and neutrons for neutron rich systems.
\item[{\tt Fig.7}] The time evolution of the N/Z ratio at the neck region for fusion
reactions of $^{40,48}$Ca+$^{90,96}$Zr. The right panel is for the
case at the energy of 10 MeV above the static Coulomb barrier and
the left panel is for the case at the energy of 5 MeV below the
static Coulomb barrier.

\item[{\tt Table.1}] The parameters used in the calculations.
\item[{\tt Table. 2}] The comparison between the static Coulomb barrier and the dynamic
barrier for reactions of $^{40,48}$Ca+$^{90,96}$Zr.
\item[{\tt Table.3}] The quantities relevant to the dynamic barrier calculated
at the time when the dynamic barrier reaches the highest value in
the fusion path.

\end{description}

\newpage

\begin{references}

\bibitem{Zag01} V. I. Zagrebaev, Phys. Rev. {\bf C64}, 034606 (2001).
\bibitem{Dia00} A. Diaz-Torres, G. G. Adamian, N. V. Antonenko, W. Scheid,
  Phys. Lett. {\bf B 481,} 228 (2000).
\bibitem{Ada00} G. G. Adamian, N. V. Antonenko, A. Diaz-Torres, W. Scheid.
  Nucl. Phys. {\bf A 671,} 233 (2000).
\bibitem{Mar01} Toshiki Maruyama, A. Bonasera, Massimo Papa, S. Chiba,
  nucl-th/0107021.
\bibitem{Ari00} Y. Aritomo, T. Wada, M. Ohta and Y. Abe, in proceedings on
 Fusion dynamics at the extremes, Dubna, 2000, edited by Yu. Oganessian and
 V. Zagrebaev ( world Scientific, Singapore 2001) P. 123;\\
 Y. Abe, C. W. Shen and G. Kosenko, in AIP conference proceedings 597 on Nonequilibrium
 and nonlinear dynamics in nuclear and other finite systems, Beijing 2001,
 edited by Z.Li, K. Wu, X. Wu, E. Zhao and F. Sakata ( American Institute of Physics)
 P. 209.
\bibitem{Tro01} M. Trotta, A. M. Stefanini, L. Corradi and et. al., Phys. Rev.
 {\bf C 65}, 011601 (R) (2001).
\bibitem{Tim98} H.Timmers, D.Ackermann, S.Beghini, L.Corradi, J.H.He, G.Montagnoli,
F.Scarlassara, A.M.Stefanini, N.Rowley, Nucl. Phys. {\bf A633}, 421 (1998).
\bibitem{Oga01} Yu. Oganessian, et. al., Phys. Rev. {\bf 63}, 011301 (R) (2001).
\bibitem{Bec85} M.Beckerman, Phys. Rep. {\bf 129},145 (1985); Rep. Prog. Phys.
 {\bf 51}, 1047 (1988).
\bibitem{Van92}R.Vandenbosch, Annu. Rev. Nucl. Part. Sci. {\bf 44}, 447 (1992).
\bibitem{Rei94} W.Reisdorf, J. Phys. {\bf G20}, 1297 (1994).
\bibitem{Bal98}A.B.Balantekin and N.Takigawa, Rev. Mod. Phys. {\bf 70}, 77 (1998).
\bibitem{Pro97}Proceedings of the International Workshop on "Heavy Ion Collisions
 at Near Barrier Energies", J. Phys. {\bf G23},1157 (1997).
\bibitem{Hag97}K.Hagino, N.Takigawa, M.Dasgupta, D.J.Hinde, and J.R.Leigh,
 Phys. Rev. Lett. {\bf 79}, 2014 (1997);
K.Hagino, N.Takigawa, and S.Kuyucak, ibid. {\bf 79}, 2943 (1997).
\bibitem{Kon00}V.N.Kondratyev, A.Bonasera, and A.Iwamoto, Phys. Rev.
 {\bf C61}, 044613 (2000).
\bibitem{Wang02} Ning Wang, Zhuxia Li, Xizhen Wu, Phys. Rev.  {\bf C 65}, 064608 (2002).
\bibitem{Hart89} Ch.Hartnack, Zhuxia Li, L.Neise, G.Peilert, A.Rosenhauser, H.Sorge,
 J.Aichelin, H.Stoecker and W.Greiner, Nucl. Phys. {\bf A495}, 303 (1989);\\
 Zhuxia Li, Ch.Hartnack, H.Stoecker and W.Greiner, Phys. Rev. {\bf C 40}, 824 (1991).
\bibitem{Ai91} J.Aichelin, Phys. Rep. {\bf 202} 233 (1991), and references therein.
\bibitem{Hart98} Ch.Hartnack, Rajeev K.Puri, J.Aichelin, Eur. Phys. J. {\bf A1},
 151 (1998).
\bibitem{Ono92}  A.Ono, H.Horiuchi, Toshiki Maruyama, and A.Ohnishi, Phys. Rev.
 Lett. {\bf 68}, 2898  (1992); Y.Kanada-En'yo and H.Horiuchi, Phys. Rev.
{\bf C52},647 (1995); Y.Kanada-En'yo, H.Horiuchi, and A. Dot\'e, Phys. Rev.
{\bf C60}, 064304 (1999).
\bibitem{Fel00} H.Feldmeier, J.Schnack, Rev.Mod. Phys.{\bf
72},655(2000).
\bibitem{Fel97} H.Feldmeier, J.Schnack, Prog. Part. Nucl. Phys. {\bf 39},
 392 (1997) and references therein.
\bibitem{Dav88} David H.Boal, James N.Glosli, Phys. Rev. {\bf C37}, 91 (1988),
and {\bf C38}, 2621 (1988).
\bibitem{Vau72} D.Vautherin, D.M.Brink, Phys. Rev. {\bf C5},626 (1972).
\bibitem{Mol95} P. M\"oller, J.R. Nix, W.D. Myers and W.J. Swiatescki, Atomic Data and Nuclear Data
{\bf59},185(1995).
\bibitem{Pap01} Massimo Papa, Toshiki Maruyama, and Aldo Bonasera, Phys. Rev.
 {\bf C 64}, 024612 (2001).
\bibitem{Li01} Qingfeng Li, Zhuxia Li, Phys. Rev. {\bf C64}, 064612 (2001).
\bibitem{Ham96} I.Hamamoto, H.Sagawa, X.Z.Zhang, Phys.Rev.{\bf C53},765 (1996).
\bibitem{Ham99} I.Hamamoto, H.Sagawa, X.Z.Zhang, Nucl.Phys.{\bf 648},203. (1999).
\bibitem{Pre75} M.A.Preston and R.K.Bhaduri, Structure of the
Nucleus, Addison-Wesley Publishing Co., Inc., Reading, Mass.
(1975) P.10-14.
\bibitem{Maru98} T.Maruyama, K.Niita, et al. Phys. Rev, {\bf C57}, 655  (1998).
\end{references}
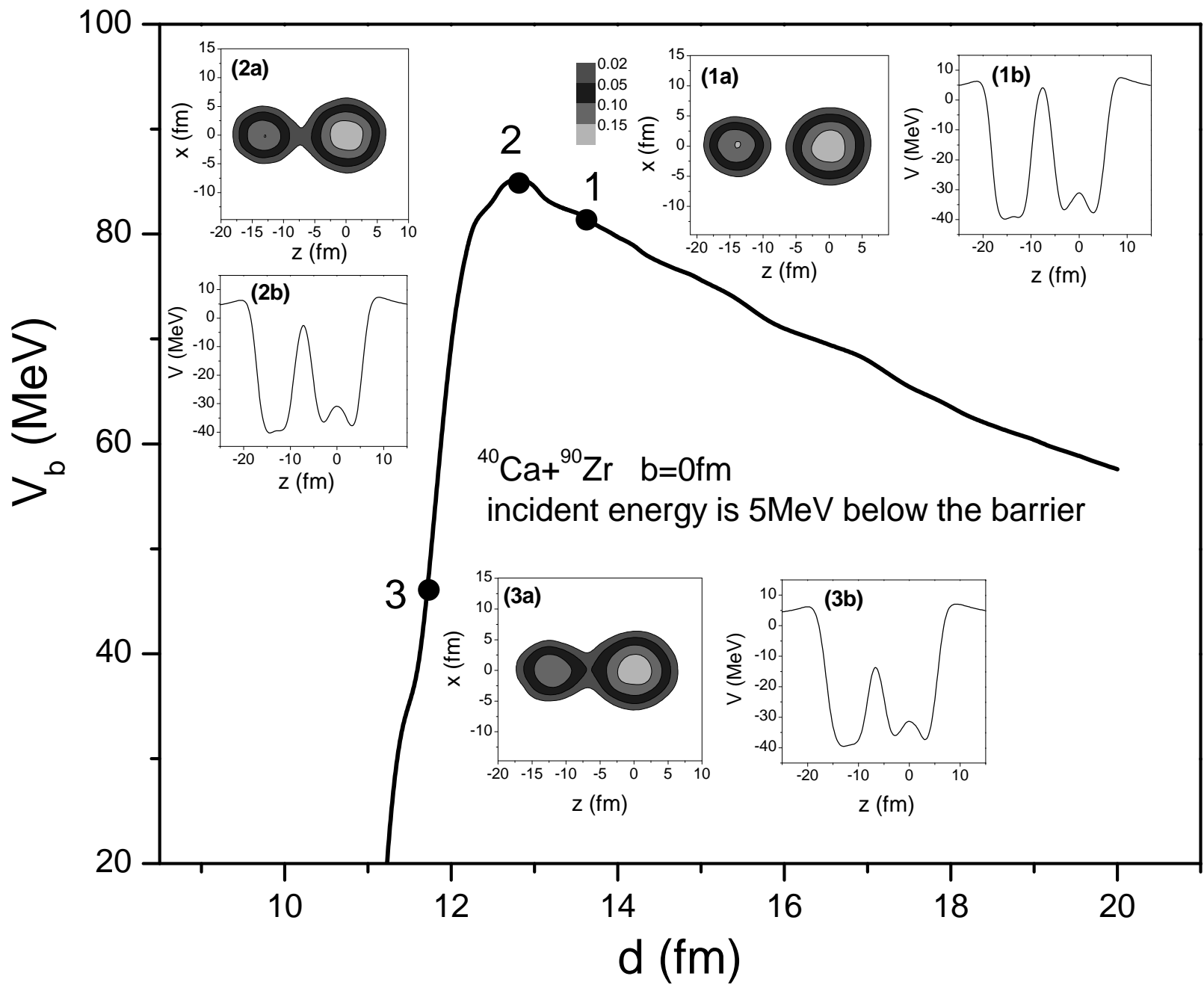
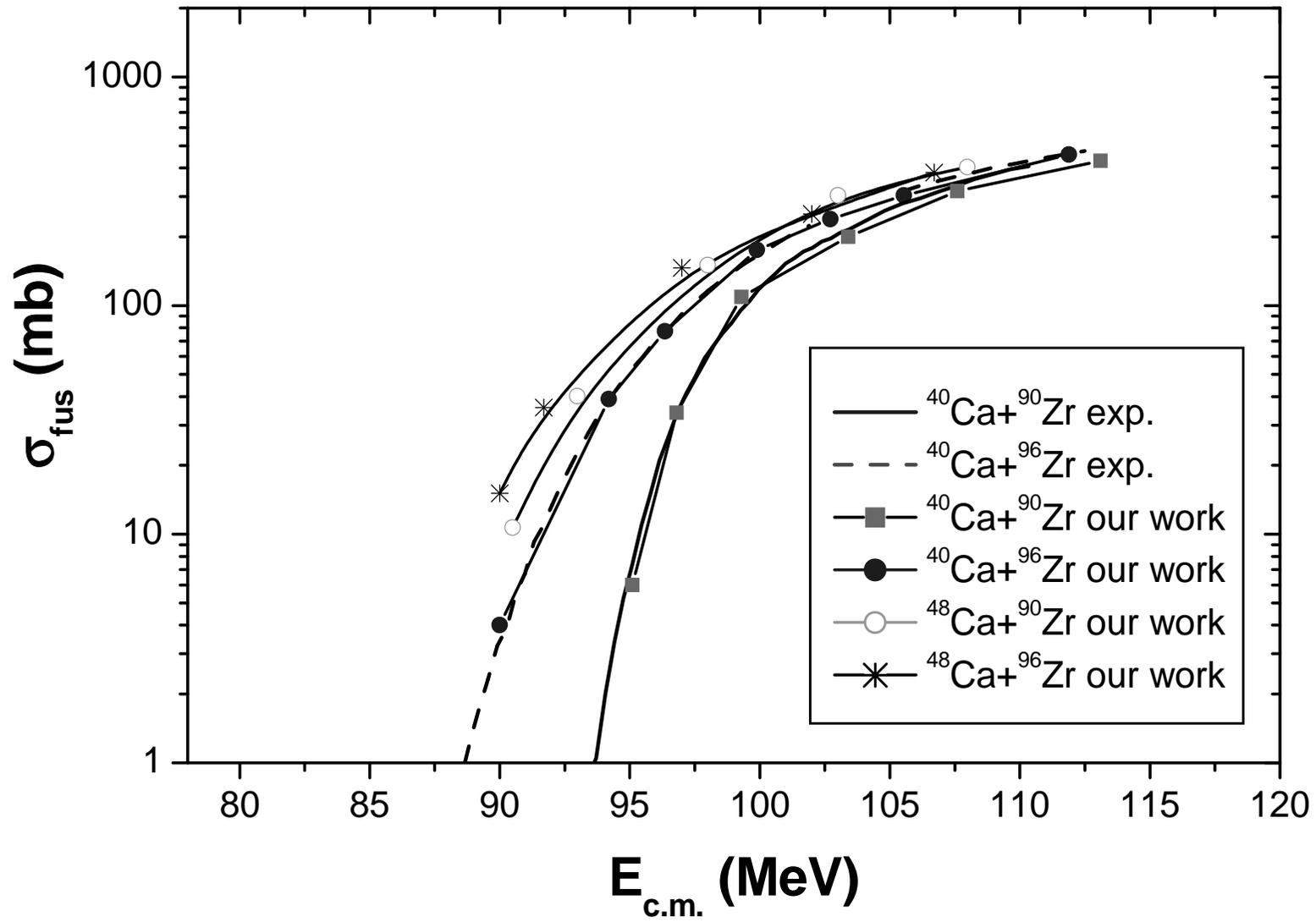
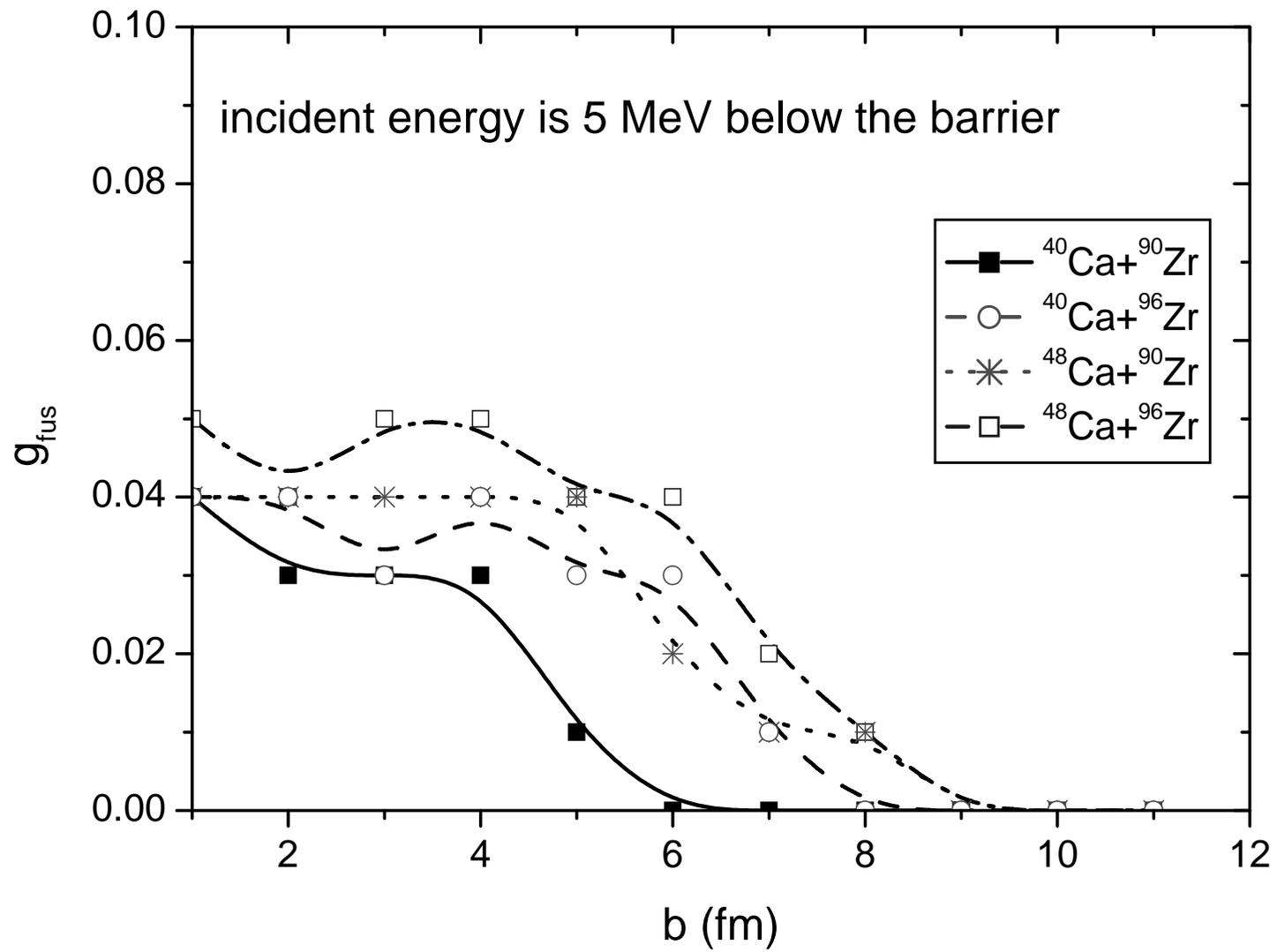
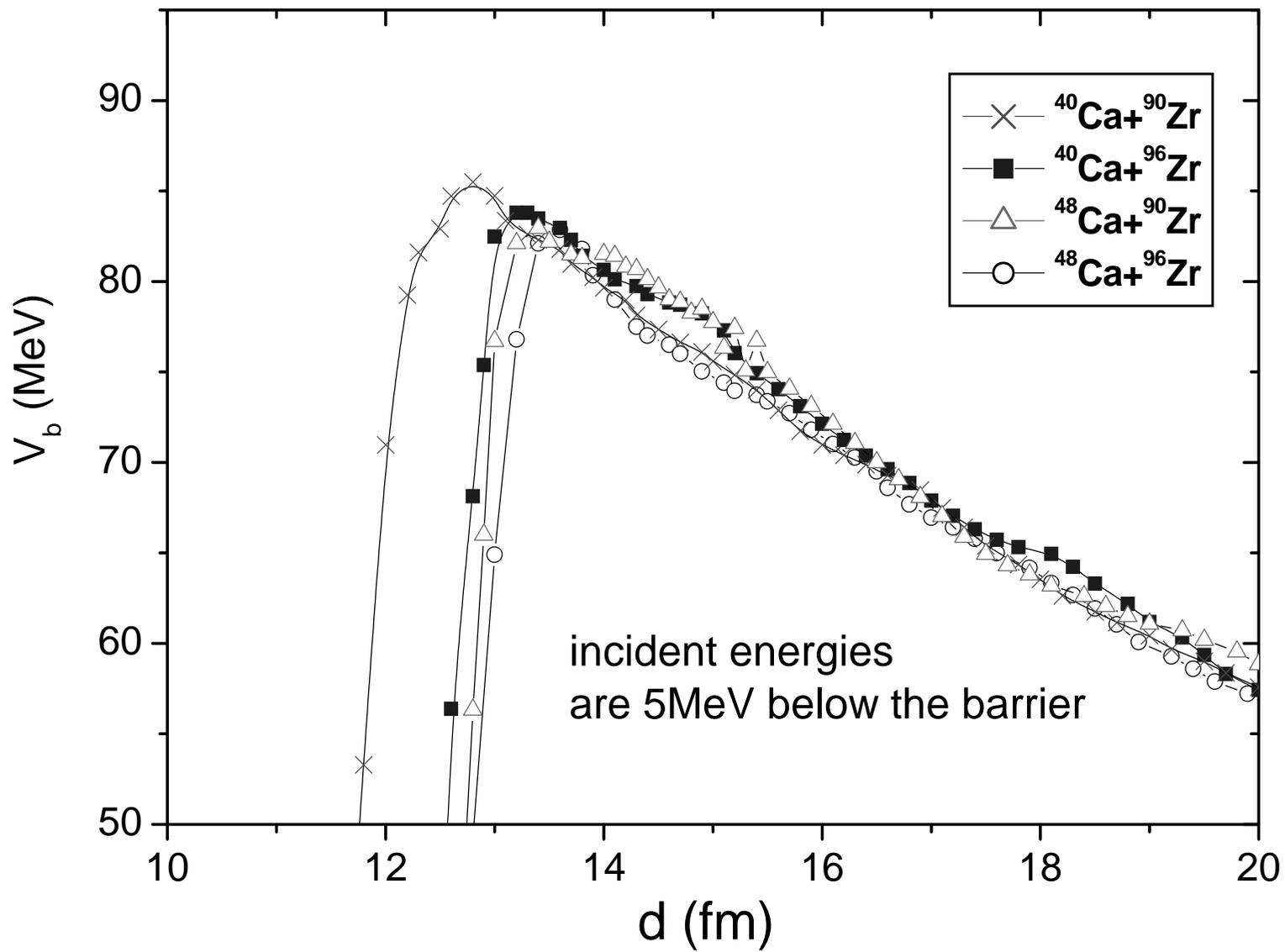
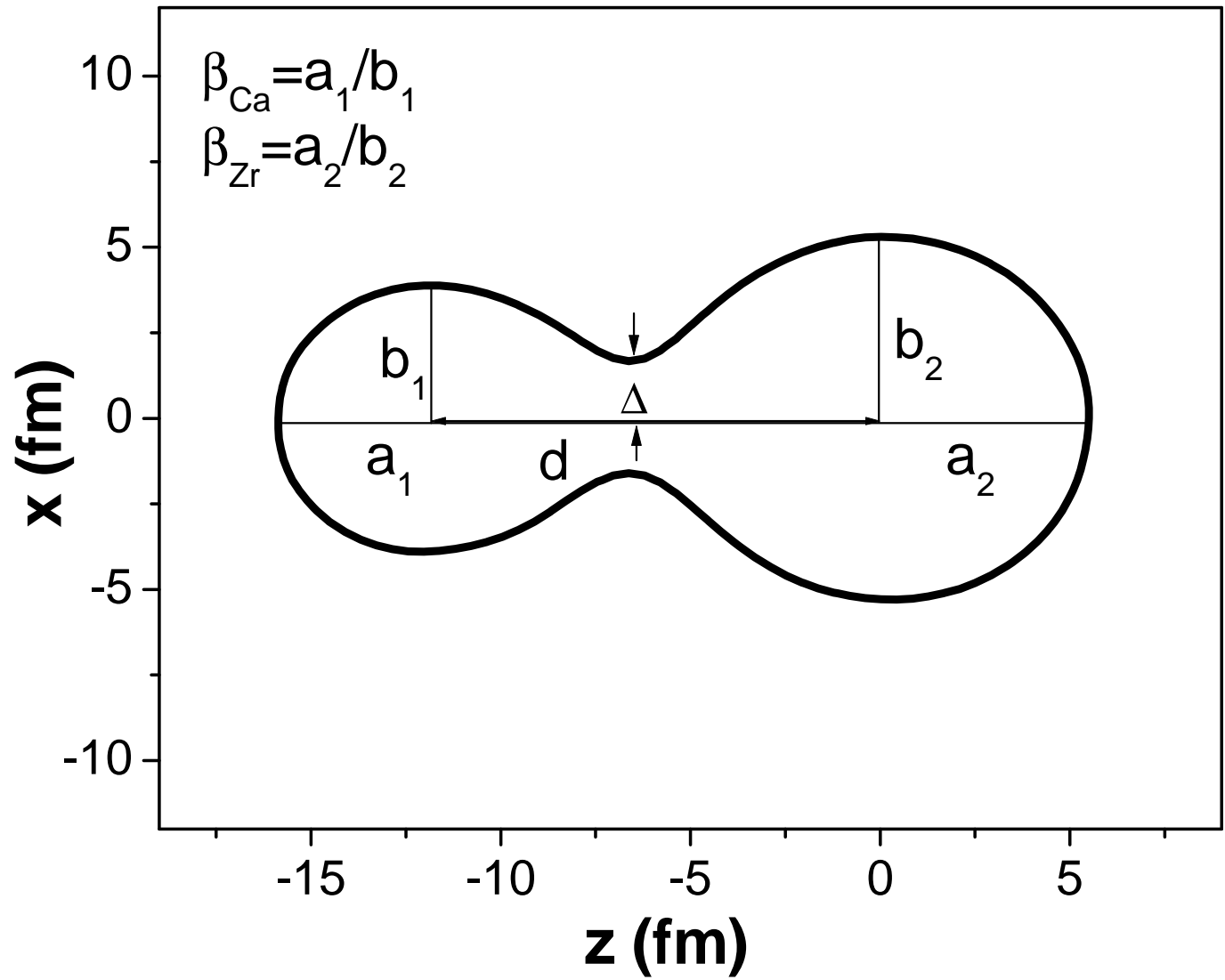
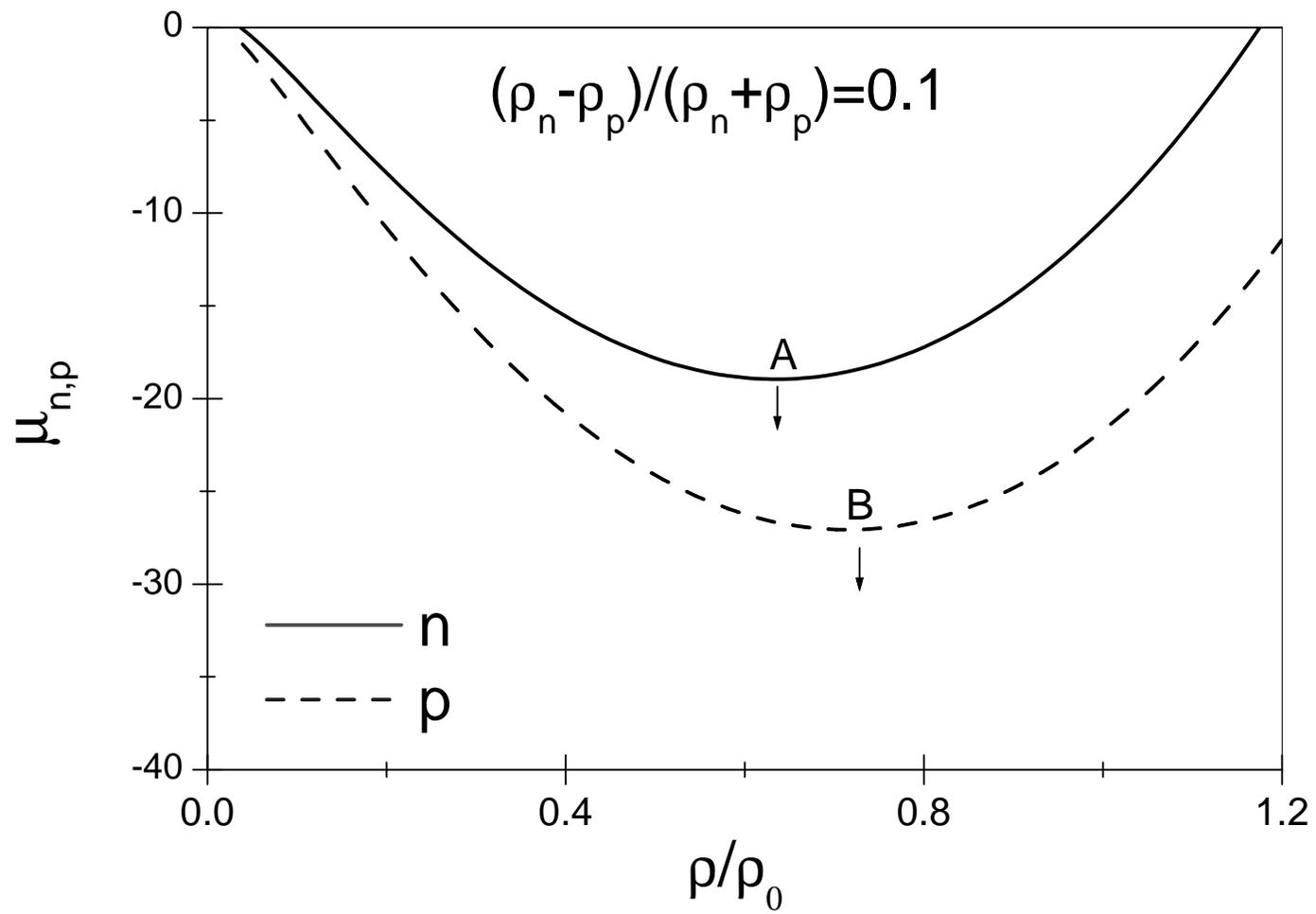
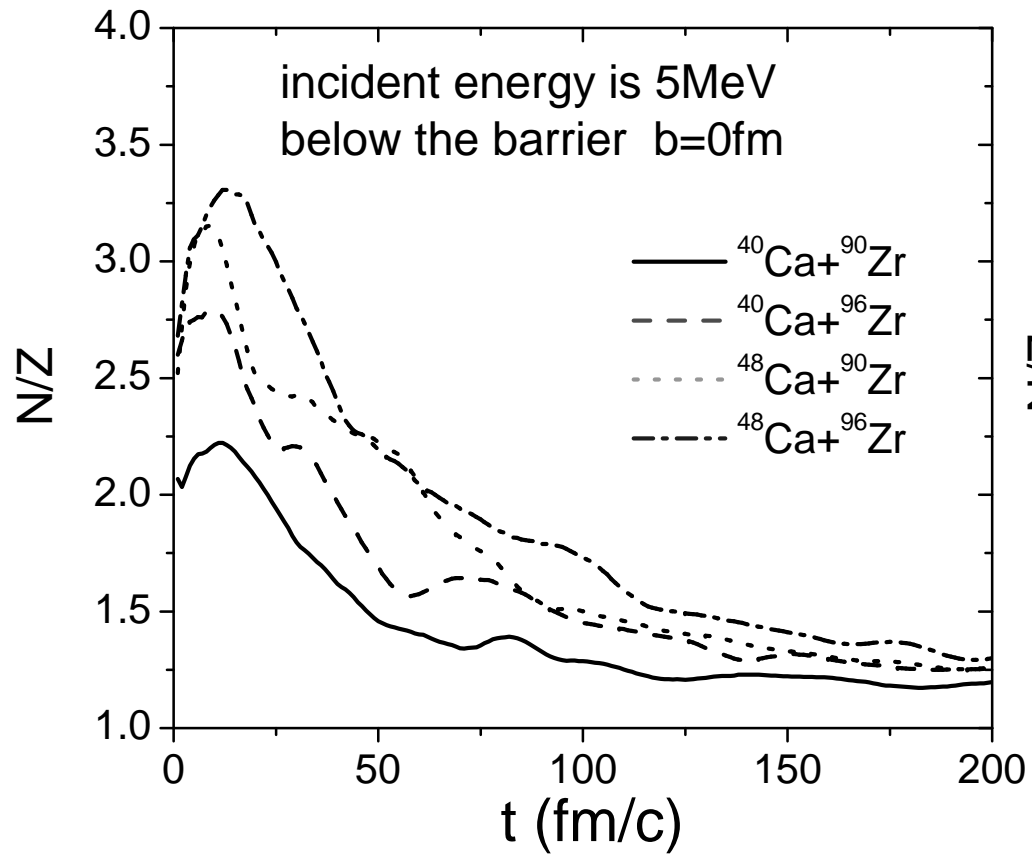
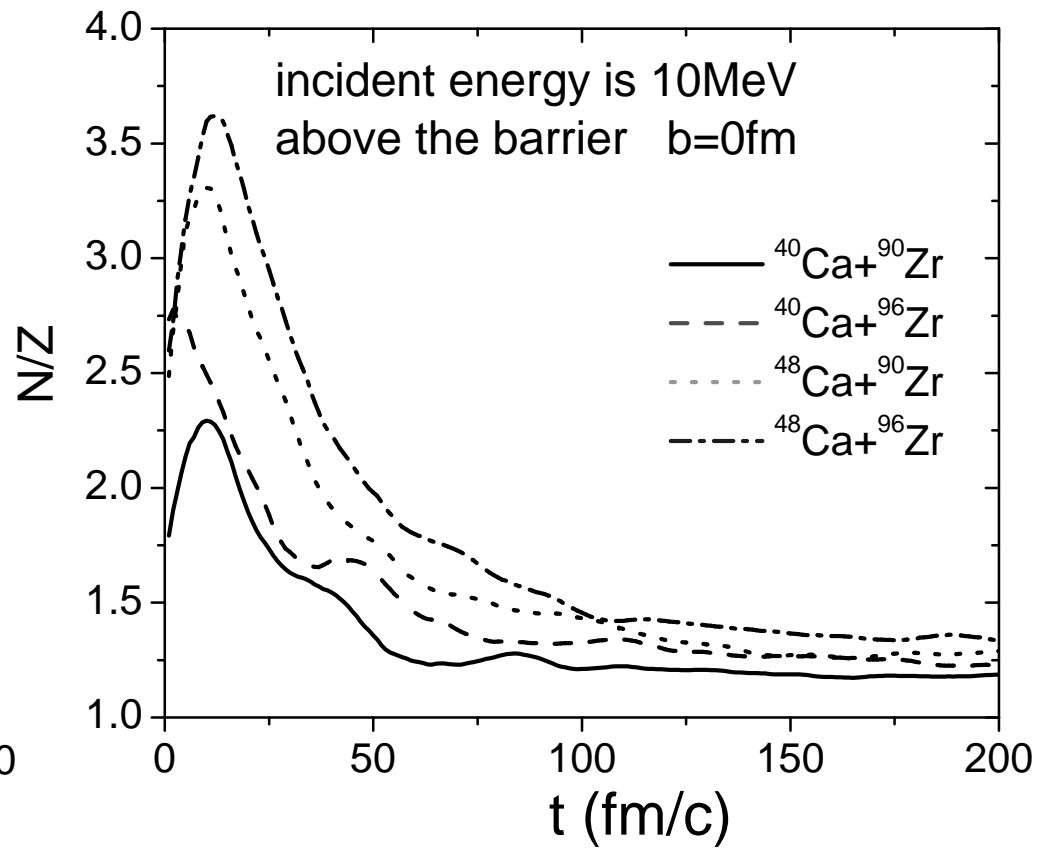
\end{document}